\documentclass[pdflatex,sn-mathphys-num]{sn-jnl}
 
\usepackage{graphicx}
\usepackage{dcolumn}
\usepackage{bm}
\usepackage{graphicx}
\usepackage[svgnames,dvipsnames]{xcolor}
\usepackage{mathtools}
\usepackage{braket}
\usepackage{qcircuit}
\usepackage{tikz}
\usepackage{amsmath}
\usepackage{amsfonts}

\definecolor{NewBlue}{rgb}{0.1, 0.1, 0.7}
\definecolor{NewRed}{rgb}{0.7, 0.1, 0.1}
\usepackage{cleveref}

\newcommand{\defeq}{\vcentcolon=}
\newcommand{\bigo}{\mathcal{O}}

\begin{document}


\title{Gradient-Based Optimization Enables Scalable Fault-Tolerant Quantum Gate Synthesis}

\author*[1]{\fnm{Marc Grau}\sur{Davis}}\email{mgd@mit.edu}
\author[2]{\fnm{Ed}\sur{Younis}}
\author[2]{\fnm{Mathias}\sur{Weiden}}
\author[1]{\fnm{Hyeongrak}\sur{Choi}}
\author[1]{\fnm{Dirk}\sur{Englund}}

\affil[1]{\orgdiv{Research Laboratory of Electronics}, \orgname{Massachusetts Institute of Technology}, \orgaddress{\street{32 Vassar St.}, \city{Cambridge}, \postcode{02139}, \state{Massachusetts}}}
\affil[2]{\orgdiv{Applied Mathematics and Computational Research Division}, \orgname{Lawrence Berkeley National Laboratory}, \orgaddress{\city{Berkeley}, \postcode{94720}, \state{California}}}

\date{\today}

\abstract{
\unboldmath
We develop a numerical approach to optimizing a quantum circuit over a discrete gate set by using gradient descent to tune $R_Z$ angles to a solution where many $R_Z$ gates can be rounded to a Clifford or $T$ gate without introducing further synthesis error. Through benchmarking, we show that our approach is able to match state-of-the-art techniques for multi-qubit Clifford + $T$ circuit synthesis, and is able to find circuits for unitaries that previous approaches fail to find solutions for. Our approach can adapt to any discrete gate set, which we demonstrate using the Clifford + $T$ + $\sqrt T$ gate set.
}

\maketitle

\section{Introduction}

Quantum computing holds the promise of significant computational advantages, from polynomial to exponential, over classical systems. Despite this potential, current quantum computers are unable to execute quantum circuits with a sufficiently large number of qubits and gates to achieve practical utility \cite{nisqbackground}. One pivotal element for scalable quantum computation is quantum circuit optimization, which is the process of rewriting a circuit to minimize resource cost.

In this context, it is crucial to distinguish between Noisy Intermediate-Scale Quantum (NISQ) computing and Fault-Tolerant (FT) quantum computing, as each has its own set of performance metrics. In NISQ architectures, the primary figure of merit is the number of two-qubit gates such as $CNOT$, $CZ$, or $iSWAP$, because the susceptibility of these gates to hardware error typically becomes the limiting factor in circuit length.  Synthesis minimizing the $CNOT$ count is thus essential for the reliability and efficiency of algorithms on NISQ devices. In contrast, in FT quantum computing, the main resource overhead comes from magic state preparation for non-Clifford gates, such as the $T$ gate. In this context, common two-qubit gates such as the $CNOT$ gate are Clifford gates and can be performed transversally on error-corrected qubits, and therefore are less of a concern. This divergence in performance metrics underscores the necessity for specialized optimization techniques, such as those targeting $T$, and by extension, $R_Z$ gate reduction.

Quantum gate synthesis is the process of taking a unitary matrix (or a similar gate description) and decomposing it into a quantum circuit. This process faces the inevitable scaling issue that unitaries scale exponentially with the number of qubits involved. Despite this scaling issue, unitary synthesis remains an essential tool in the quantum compilation stack, where it sees applications in transpilation between different gate sets \cite{ion_numopt, french, cosed}, discovery of new patterns to apply in circuit optimization \cite{constant}, as a way to generate multiple compilations of a circuit for averaging out errors \cite{randomized_compiling, hastings2016turning}, and as an optimization step in itself by computing the unitary for a circuit and synthesizing a more efficient version \cite{qgo, ml_circopt}. Typically the scaling problem is mitigated by focusing on small gates, either from small unitaries or as subcircuits taken from a larger circuit.

There many different approaches to quantum gate synthesis, including recursive matrix decompositions \cite{qsd, CSD, solovay-kitaev}, numerical approaches \cite{qsearch, LEAP, ion_numopt, french, best_approx}, and some more specialized approaches such as optimized decomposition of the two qubit case \cite{kak, kak2}.

Synthesis techniques intended for NISQ computing take advantage of continuously parameterized gates that can be implemented in hardware, such as $R_Z$ and the Mølmer-Sørensen gate. These continuous parameters are what enable numerical approaches to NISQ synthesis, which use the gradient of a cost function with continuous parameters to guide the search for a numerical solution. In the context of FT quantum computing, the set of allowed gates is restricted to a discrete gate set, typically consisting of the Clifford gates ($X$, $Y$, $Z$, $H$, $S$, $CNOT$, and any combination of these gates), along with at least one non-Clifford gate (the most common choice is $T$). This discrete gate set prevents numerical synthesis from being applied to FT quantum synthesis.

In FT quantum computing, we consider the Clifford+$T$ gateset with a primary focus on the minimization of $T$ gates.  This gateset can only exactly implement unitaries in the ring $\mathbb{Z}[i,\frac{1}{\sqrt{2}}]$ \cite{gidney2021factor, exact_multi}. Other unitaries must be approximated by a unitary in that ring that is ``close'' by some error metric.

Many current approaches to Clifford+$T$ synthesis involve a combinatorial search over possible circuits \cite{meet_in_middle, synthetiq, t_any}.  Some approaches instead focus on the approximation of single-qubit unitaries \cite{gridsynth, solovay-kitaev, kliuchnikov2022shorter, rus_clift} or very specific sets of multi-qubit unitaries \cite{diagonal_synth, another_toffoli}.

A common approach for approximating arbitrary multi-qubit unitaries with the Clifford+$T$ gateset involves first finding an (exact or approximate) implementation of the target unitary in a continuous gateset (such as Clifford+$R_Z$), and then using single-qubit approximation technqiues to find Clifford+$T$ approximations for the non-Clifford gates in the circuit \cite{quantumsim, diagonal_synth}.

In this work, we improve this approach to arbitrary multi-qubit unitary approximation: given a Clifford+$R_Z$ circuit, we use numerical optimization to minimize the $T$-count of the circuit after single-qubit gate approximation is applied.  In this way, our algorithm serves as part of the transpilation process for converting a Clifford+$R_Z$ circuit generated by another technique, such as numerical synthesis \cite{qsearch, LEAP, french, diagonal_synth} or construction from an algorithm \cite{quantumsim}, into a Clifford+$T$ circuit.

When this conversion from Clifford+$R_Z$ to Clifford+$T$ is performed, each $R_Z$ gate becomes approximated by a long circuit consisting of 10s to 100s of $T$ gates, depending on the synthesis error allowed for the approximation \cite{gridsynth, solovay-kitaev, kliuchnikov2022shorter, rus_clift}.  In this way, $R_Z$ gates become the biggest contributing factor to the final $T$-count of the circuit. The length of these sequences is independent of the value of the $R_Z$ angle, except when the gate is within the error threshold of a Clifford or $T$ gate.

Our approach uses numerical optimization to identify $R_Z$ gates that can be rounded to a Clifford or $T$ gate such that they do not need to be approximated using a single-qubit synthesis technique. We also minimize the number of $T$ gates added during this process. By focusing on minimizing the biggest contributions to $T$-count, and we aim to directly tackle the most contributing factor to FT quantum circuit resource usage.



\section{Results}

\begin{table}
\centering
\begin{tabular}{|l|r||r|r||r|r|r|r|}
\hline
\textbf{Gate} & \textbf{Qubits} & \textbf{TRbO $T$} & \textbf{Synthetiq $T$} & \textbf{TRbO}\footnotemark[1] \textbf{(s)} & \textbf{Synthetiq (s)}\\
\hline
\hline
QFT & 2 & 3 & 3 & 16 & 2\\
Toffoli & 3 & 7 & 39 & 53 & 2\\
Grover & 3 & 14 & Timeout & 40 & $>$ 3600 \\
\hline
\end{tabular}
\footnotetext[1]{This runtime includes the LEAP synthesis pass that converts the unitary to a Clifford + $R_Z$ circuit before TRbO can be run.}
\vspace{1em}
\begin{tabular}{|l|r||r|r||r|r|r|r|}
\hline
\textbf{Gate} & \textbf{Qubits} & \textbf{LEAP $R_Z$} & \textbf{TRbO $R_Z$} & \textbf{LEAP (s)} & \textbf{TRbO (s)}\\
\hline
\hline
QFT & 2 & 24 & 0 & 4 & 12\\
Toffoli & 3 & 57 & 0 & 14 & 39\\
Grover & 3 & 63 & 0 & 10 & 30\\
QFT & 3 & 45 & 3 & 8 & 11\\
TFIM & 3 & 51 & 17 & 8 & 18\\
$C\sqrt{iSWAP}$ & 3 & 63 & 4 & 12 & 18 \\
Heisenberg & 3 & 93 & 33 & 14 & 121 \\
\hline
\end{tabular}

\vspace{1em}
\caption{$T$-count comparison with Synthetiq\cite{synthetiq} (top) and $R_Z$-count comparison with LEAP\cite{LEAP} as a baseline (bottom).}
\label{tab:synth}
\end{table}
We implemented the technique described in Section \ref{sec:procedure} as a Python package that we call TRbO ($T$ Reduction by Optimization). We benchmarked TRbO by using it to optimize a suite of reference circuits and compared our results to state-of-the art multi-qubit Clifford + $T$ circuit optimization techniques \cite{synthetiq, t_any, autoopt}.
\subsection{Synthesis of Small Unitaries}
To evaluate the use of TRbO in a gate synthesis workflow, we ran the LEAP synthesis tool from BQSKit \cite{LEAP, bqskit} on several common gates that are known to have exact Clifford + $T$ decompositions, such as the Toffoli gate and the 2-qubit Quantum Fourier Transform (QFT). We were able to reproduce the best-known $T$ count for these gates. We also processed several of the smaller gates from the BQSKit benchmarking set. We also ran Synthetiq \cite{synthetiq} for comparison, although it hit the time limit of 1 hour in most cases. Both Synthetiq and TRbO were configured to take the best of 10 results, with multithreading enabled, and timed for the total of all 10 runs. TRbO was run on ansatz circuits produced by LEAP, and the LEAP synthesis time is reported as well. The results are shown in Table \ref{tab:synth}.

We find that TRbO is generally slower at processing simpler circuits, especially when including LEAP synthesis time, but TRbO scales much better as the problem difficulty increases. TRbO was able to find a Clifford + $R_Z$ circuit for the ``Grover 3'' benchmark from bqskit in only 40 seconds, while Synthetiq was unable to find any solution for this unitary before it hit the timeout we set at 1 hour. Also, TRbO is capable of handling circuits where a short Cliford+$T$ circuit is impossible to find, such as the 3-qubit Quantum Fourier Transform (QFT), for which \textit{Gheorghiu et al.} reported that their tool failed to find a solution \cite{t_any}. Similarly, Synthetiq timed out on all circuits for which TRbO was unable to find a solution with no leftover $R_Z$ gates. We note that the LEAP and TRbO combination struggled at some of the harder benchmarks from Synthetiq's benchmarking set, such as $C\sqrt{iSWAP}$. We didn't run this with Synthetiq ourselves because the Synthetiq paper notes that this gate took 4 hours to synthesize. Further discussion is in Section \ref{sec:ansatz}.

\subsection{Optimization of Large Circuits}

We compare TRbO to the results published by \textit{Nam et al.} \cite{autoopt} to benchmark the use of TRbO as an optimization tool for large circuits. We used the ``QuickPartitioner'' from BQSKit \cite{bqskit} to split the large circuits into smaller circuits, and used TRbO to reduce the $R_Z$ count in those smaller circuits. When we ran TRbO on the already-optimized outputs from \textit{Nam et al.}, we found further optimization opportunities. We focused on circuits with $R_Z$ gates left in the output, which were either QFT or QFTAdd circuits. Of those we saw no improvement in the QFT circuits but a small improvement in the QFTAdd circuits. The results from these QFTAdd circuits are shown in Table \ref{tab:quipper}. We were unable to do further analysis or comparison due to the lack of code availability from \textit{Nam et al.}. We believe that the results of TRbO in this context can be improved with further research into circuit partitioning, and discuss this in Section \ref{sec:ansatz}.

\begin{table}[ht!]
\centering
\begin{tabular}{|l|r||r|r||r|}
\hline
\textbf{Gate Name} & \textbf{Qubits} & \textbf{Prev. $R_Z$} & \textbf{TRbO $R_Z$} & \textbf{Time (s)} \\
\hline
\hline
QFTAdd8 & 16 & 122 & 112 & 10 \\
QFTAdd16 & 32 & 420 & 402 & 23 \\
QFTAdd32 & 64 & 1076 & 1042 & 74 \\
QFTAdd64 & 128 & 2388 & 2322 & 159 \\
QFTAdd128 & 256 & 5012 & 4882 & 747 \\
QFTAdd256 & 512 & 10260 & 10002 & 550 \\
QFTAdd512 & 1024 & 20756 & 20242 & 1294 \\
QFTAdd1024 & 2048 & 41748 & 40722 & 3931 \\
\hline
\end{tabular}
\vspace{1em}
\caption{Results from running TRbO on already-optimized outputs from Nam et al. \cite{autoopt}}
\label{tab:quipper}
\end{table}

\subsection{Beyond \textit{T} Gates} \label{sec:sqrtt}

We have chosen to focus on $T$ because it is a common choice of non-Clifford gate. However, the minimization problem in equation \ref{eq:min_problem} is not limited to solving for Clifford and $T$ gates.  We can set $D$ to be any discrete set of angles that we wish, allowing us to consider other discrete gate sets. In principle, this technique can be modified to convert any continuous gate set to any discrete gate set. For example, the 3-qubit quantum Fourier transform typically requires 3 $R_Z$ gates to be synthesized, but by employing $\sqrt{T}$, which has been studied as a tool for more efficient gate approximation sequences \cite{kliuchnikov2022shorter, sqrttuse}, we are able to find a circuit with no $R_Z$ gates, 5 $\sqrt{T}$ gates, and 9 $T$ gates.

\section{Discussion}

In this work, we developed a gradient descent powered approach to quantum circuit optimization with a discrete target gate set. We derived a numerically minimizable bound on the synthesis error from rounding $R_Z$ angles to the nearest value from a discrete set, and use this to discover sets of parameters for the $R_Z$ angles in a circuit that allow many $R_Z$ gates to be safely rounded to a Clifford gate or a single $T$ gate. Adding a binary search for the maximum number of $R_Z$ gates that can be rounded this way, we have created a practical and performant approach to $T$ count minimization that avoids the combinatorial search found in many other approaches \cite{synthetiq, advanced_synth, t_any}.

Our approach can easily be adapted to other discrete gate sets, as we demonstrate by working with the Clifford + $T$ + $\sqrt{T}$ gate set in Section \ref{sec:sqrtt}. Future work might include extending this approach beyond the Clifford hierarchy, such as considering approaches to creating more general classes of non-Clifford magic states \cite{choi2023fault}.

We expect that future improvements in circuit partitioning techniques will improve the performance and result quality of scaling techniques such as ours to circuits with large numbers of qubits. We also expect that approaches such as ours can be used to discover patterns and identities to be used in template based optimization techniques such as the results from \text{Nam et al.}\cite{autoopt}. Numerical synthesis techniques have already been used to discover new approaches to circuit generation in the NISQ space \cite{constant}.

\section{Methods}
\label{sec:procedure}
\subsection{Quantum Process Infidelity}
As our metric of synthesis error, we use the process infidelity resulting from replacing the target unitary with our implemented one.

\begin{equation} \label{eq:dist_func}
d(U_1,U_2) = \sqrt{1-\frac{\lvert\lvert \operatorname{Tr}(U_1^\dagger U_2)\rvert\rvert^2}{\operatorname{dim}(U_1)^2}}
\end{equation}

This distance function and related ones are commonly used as error metrics in synthesis papers because of their statistical meaning and relatively low computational complexity, including the ease of computing the derivative \cite{qsearch, LEAP, quest, t_any, synthetiq}.

Whenever we mention ``synthesis error'' in this paper, we are referring to this error metric.

An important property of this error metric is that we can compute an upper bound on the error of a circuit consisting of multiple gates by adding the error of each component gate \cite{quest}.

\begin{equation} \label{eq:add_error}
    d(U_1V_1, U_2V_2) \leq d(U_1,U_2) + d(V_1,V_2)
\end{equation}

To prove this property, first we note that this metric is not affected by using the Kronecker product with the identity to extend a unitary to act on more qubits.

\begin{equation}\label{eq:dist_eq}
d(U_{1}\!\otimes I,\; U_{2}\!\otimes I)=d(U_{1},\,U_{2}) \\
\end{equation}
\begin{center}
Which can be derived from the following properties which hold for any same-sized unitaries $U_1$ and $U_2$, and can be easily verified:
\end{center}
\begin{equation}
\begin{aligned}
I &\defeq
\begin{pmatrix}
1 & 0\\
0 & 1
\end{pmatrix}\\[4pt]
(U_{1}^{\dagger}\!\otimes I)\,(U_{2}\!\otimes I)
&= (U_{1}^{\dagger}U_{2})\!\otimes I\\[4pt]
\operatorname{Tr}(U\!\otimes I)
&= 2\,\operatorname{Tr}(U)\\[4pt]
\dim(U\!\otimes I)
&= 2\,\dim(U)
\end{aligned}
\end{equation}
\begin{center}
\end{center}

Next we note the following property derived by Bo-Ying Wang and  Fuzhen Zhang \cite{trace_inequality}
\begin{equation} \label{eq:trace_inequality}
\begin{aligned}
\resizebox{0.9\hsize}{!}{$\sqrt{1-\frac{\lvert\lvert \operatorname{Tr}(U_1U_2)\rvert\rvert^2}{n^2}} \leq \sqrt{1-\frac{\lvert\lvert \operatorname{Tr}(U_1)\rvert\rvert^2}{n^2}} + \sqrt{1-\frac{\lvert\lvert \operatorname{Tr}(U_2)\rvert\rvert^2}{n^2}}$}\\
\resizebox{0.9\hsize}{!}{$\textit{For any unitaries $U_1$, $U_2$ such that }
\operatorname{dim}(U_1)=\operatorname{dim}(U_2)=n$}
\end{aligned}
\end{equation}

Equation \ref{eq:trace_inequality} allows us to compute a bound on the total synthesis error by simply adding per-gate errors, and Equation \ref{eq:dist_eq} allows us to do this even when these gates are not acting on the same set of qubits.

\subsection{The Synthesis Error Cost of Rounding \textit{R\textsubscript{Z}} Gates}

If we replace an $R_Z$ gate with another $R_Z$ gate with an angle that is different by $\theta$, the synthesis error of this replacement is $\sqrt{\frac{1}{2}(1-\cos(\theta))}$.  Using the small angle approximation $\cos(\theta)\approx 1-\frac{1}{2}\theta^2$, we can simplify this error cost to $\frac{1}{2}\lvert\theta\rvert$ as long as the angle $\theta$ is small.  When $\theta$ is large, it is sufficient for our purposes for our computation to yield a value that is large compared to the success threshold we place on the error of the overall circuit, such that we can reject this replacement.

\begin{figure*}
\includegraphics[width=\textwidth]{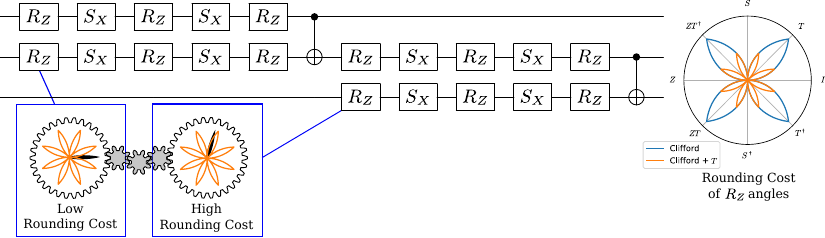}
\caption{Polar plot of the gate rounding cost function, and depiction of how distant $R_Z$ angles can effect each other during optimization.}\label{fig:flowers} \it \footnotesize
\end{figure*}

If we define a set of ``desired'' angles, such as the set of $R_Z$ angles corresponding to Clifford gates:
\begin{equation}
    D \defeq \{\text{The set of ``desired'' $R_Z$ angles.}\}
\end{equation}
If we replace the angle $\theta$ of an $R_Z$ gate with $\theta^\prime$, such that $\theta^\prime \in D$, the error added by this operation can be described by Equation \ref{eq:angle_error}. This results in a triangle wave cost function, plotted in the polar plot on the right of Figure \ref{fig:flowers}.

\begin{equation} \label{eq:angle_error}
\epsilon(\theta) = \min_{\phi \in D}\frac{\lvert\theta - \phi \rvert}{2}
\end{equation}

Now if we repeat this process for many gates in a circuit, we can compute an upper bound on the synthesis error after performing this ``rounding'' procedure using equation \ref{eq:add_error} by adding the errors together.
\begin{equation}
d(U_\text{target},U(\vec{\theta}^\prime) \leq d(U_\text{target},U(\vec{\theta})) + \sum{\epsilon(\vec{\theta})}
\end{equation}

\subsection{Optimization Problems for Desired $R_Z$ Angles}

If we decide to round $N$ of the $M$ $R_Z$ gates in the circuit ($N \leq M$), and wish to find a solution that minimizes the synthesis error after performing this rounding, we can write the error from performing this operation as:

\begin{equation}
\vec{\epsilon}(\vec{\theta}) \defeq \{\epsilon(\theta_i) \forall \theta_i \in \vec{\theta} \text{, sorted by value of }\epsilon\}
\end{equation}
Finding a set of $R_Z$ angles that minimize this error is now a continuous valued numerical optimization problem:
\begin{equation} \label{eq:min_problem}
    \min d(U_\text{target},U(\vec{\theta})) + \frac{1}{2}\sum_{i=1}^N{\vec{\epsilon}_i(\vec{\theta})}
\end{equation}

This cost function forces $N$ of the $R_Z$ gates towards one of the desired gates in $D$, while allowing the remaining $R_Z$ gates to respond freely as needed to keep the implemented unitary close to the target. This behavior is depicted as distant $R_Z$ gates connected by gears in Figure \ref{fig:flowers}.

To find a gate rounding solution that minimizes the number of remaining $R_Z$ gates, we want to maximize $N$, while keeping the cost of rounding $N$ gates below some success threshold $T$.

\begin{equation} \label{eq:max_N}
\begin {aligned}
    \max N \\
    \text{s.t. } d(U_\text{target},U(\vec{\theta})) + \frac{1}{2}\sum_{i=1}^N{\vec{\epsilon}_i(\vec{\theta})} \leq T
\end{aligned}
\end{equation}

This is a single variable integer optimization problem, so we can binary search over the range of possible values of $N$, from $0$ to the $M$, the total number of $R_Z$ gates in the circuit.  For each value of $N$, we can solve the minimization problem in equation \ref{eq:min_problem}, and check that its solution is below the threshold.  This procedure will allow us to find a solution for both $N$ and $\vec{\theta}$ that allows us to round as many $R_Z$ gates from the circuit as possible to gates in the ``desired'' set $D$, while keeping the synthesis error low below the threshold $T$.

\subsection{Minimizing \textit{T} Count}

To minimize the $T$-count of a Clifford+$R_Z$ circuit, we first use the set of $R_Z$ angles corresponding to both Clifford and $T$ gates as our set desired angle set $D$:
\begin{equation}
    D = \{\theta \frac\pi4 \forall \theta \in \mathbb{Z} \}
\end{equation}

When we finish solving the binary search problem, we get $N_{\text{Cliff}+T}$, the total number of $R_Z$ gates that can be rounded to either Clifford or $T$ gates without adding too much synthesis error.

Next, we do another binary search to find $N_{\text{Cliff}} \leq N_{\text{Cliff}+T}$, the maximum number of $R_Z$ gates that can be rounded to Clifford gates such that only $M - N_{\text{Cliff}+T}$ $R_Z$ gates are leftover to be approximated by single-qubit synthesis.  We solve for $N_{\text{Cliff}}$ by first solving the numerical optimization problem in equation \ref{eq:min_problem}, for rounding $N_{\text{Cliff}}$ gates to $D = \{\theta \frac\pi2 \forall \theta \in \mathbb{Z} \}$ (the set of $R_Z$ angles corresponding to Clifford gates). Then we perform gate rounding and put the remaining $R_Z$ gates into another optimization problem, this time for rounding $N_T = N_{\text{Cliff}+T} - N_{\text{Cliff}}$ $R_Z$ gates to $T$ gates. We ensure that this optimization problem achieves a solution below the threshold $T$ before accepting $N_{\text{Cliff}}$.

By following this procedure, we ensure that we round as many gates as possible to Clifford gates, and then round as many remaining gates as possible to $T$ gates, while minimizing the number of $R_Z$ gates leftover after this process.

\subsection{Selection of Ansatz Circuit} \label{sec:ansatz}

For this work we implemented the procedure for modifying $R_Z$ angles and replacing them with Clifford + $T$ circuits, bit we rely on other sources for the initial Clifford + $R_Z$ ansatz circuits. In most cases we used the existing $CNOT$ structure of the circuit while converting all single-qubit gates using $R_ZS_XR_ZS_XR_Z$ or similar decompositions. Where this structure couldn't be applied (such as when the initial circuit contained non-Clifford multi-qubit gates), we used numerical synthesis to find a Clifford+$R_Z$ circuit \cite{LEAP}. The ansatz circuit derived in this way is designed to minimize $CNOT$ count, and this does not always allow a fully minimized $T$ count. For example, when synthesizing a controlled-$\sqrt{iSWAP}$ gate, the combination of LEAP and TRbO finds a solution with 6 CNOTs and 4 $R_Z$ gates. However, when using the $CNOT$ structure from one of the results published for Synthetiq \cite{synthetiq}, TRbO finds a result with 20 CNOTs, 10 $T$ gates, and 0 $R_Z$ gates, matching the best known solution.

\subsection{Optimization of Large Circuits}
Solving equation \ref{eq:max_N} requires solving equation \ref{eq:min_problem} $\bigo(\log l)$ times, where $l$ is the length of the circuit (the number of gates).  Each of those solutions is a continuous numerical optimization problem with a cost function involving $d(U_\text{target}, U(\vec{\theta}))$, and those unitaries will have dimension $2^n$ where $n$ is the number of qubits involved in the circuit.  This adds an exponential factor to the overall runtime, allowing us to only optimize circuits with a small number of qubits.  This is a common limitation among synthesis algorithms.

However, it is possible to use synthesis techniques to optimize larger circuits by partitioning the circuit into smaller sub-circuits and optimizing those subcircuits independently.  This technique has been to apply numerical synthesis as an optimization technique for large quantum circuits \cite{qgo, quest}.  When partitioning circuits in this way, the exponential factor is now in the block size of the partitioning, which can be used as a tradeoff between runtime and the quality of the overall optimization.  The runtime scaling in the size of the circuit, assuming a constant block size, becomes $\bigo(l \log l)$ optimizations of a fixed size. We used the ``QuickPartitioner'' from BQSKit \cite{bqskit} in our benchmarking.

\subsection{Software Implementation}

We implemented our optimization algorithm as a plug-in ``pass'' for the quantum circuit optimization library BQSKit\cite{bqskit}. Our code is written in Python, but uses the optimized quantum circuit simulation and numerical optimization tools provided in BQSKit, which are written in Rust. We also use BQSKit's runtime system to enable multithreading. We focus on the step of minimizing $R_Z$ gates, but include a wrapper for ``gridsynth''\cite{gridsynth} as a BQSKit pass to perform the final step of converting $R_Z$ gates to single-qubit Clifford+$T$ circuits.

We found a significant benefit, from re-using successful sets of parameters between instances of solving the minimization problem in equation \ref{eq:min_problem}, even between different values of $N$ or different target sets of angles $D$. Similarly, we used a two-step optimization approach, where we first used random seeds to find multiple solutions that minimize the matrix distance from equation \ref{eq:dist_func} without any consideration of the specific $R_Z$ angles. Then, we used those solutions as the starting points for solving equation \ref{eq:min_problem}. This procedure of carefully choosing starting points using the results of other optimization problems turned out to be roughly a 30\% speedup and in some cases improved the quality of results.

\subsection{Benchmarking Setup}

All of our benchmarking was done on a Linux desktop with an AMD 5950X 16-core / 32-thread CPU and 128GB of RAM. We developed a wrapper for Synthetiq to be used as a BQSKit pass for comparison to our approach \cite{synthetiq}. We also compare to the results published by \textit{Gheorghiu et al.} \cite{t_any} and \textit{Nam et al.} \cite{autoopt}. For all benchmarking we ran TRbO with default settings. For comaprison with Synthetiq, we ran TRbO 10 times and took the best result, to mimic the default behavior of Synthetiq.

\section*{Data Availability}
All inputs used in benchmarking, a JSON file summarizing outputs, and the code needed to reproduce the results are included in the TRbO repository at: \href{https://github.com/WolfLink/trbo}{https://github.com/WolfLink/trbo} in the ``benchmarks'' folder.

\section*{Code Availability}
All code developed for this paper is available open-source under the MIT license in the TRbO repository at: \href{https://github.com/WolfLink/trbo}{https://github.com/WolfLink/trbo}

\section*{Acknowledgements}
\noindent We thank Emma Smith for fruitful discussions and expert Python advice. This material is based upon work supported by the U.S. Department of Energy, Office of Science, Office of Advanced Scientific Computing Research, Department of Energy Computational Science Graduate Fellowship under Award Number DE-SC0021110. This material is based upon work supported by the U.S. Department of Energy, Office of Science, National Quantum Information Science Research Centers, Quantum Systems Accelerator under Award No. DE-SCL0000121. This material is based upon work supported by the National Science Foundation under Award No. 2419204.

\section*{Author Contributions}
M.D. conceived the main idea, developed the Python code, and drafted the manuscript.
E.Y. and M.W. provided key insights and assisted with supporting code, and E.Y. edited the manuscript.
H.C. advised on the research direction and edited the manuscript.
D.E. supervised the project.
All authors reviewed and approved the final manuscript.

\section*{Competing Interests}
All authors declare no competing interests. 

\bibliography{references}




\end{document}